# Drug-Associated Rhabdomyolysis in the FDA Adverse Event Reporting System (FAERS): A Disproportionality Analysis

Enpu Liang

## Abstract

**Background:** Rhabdomyolysis is a serious adverse drug reaction associated with a wide range of drugs. We screened the U.S. Food and Drug Administration Adverse Event Reporting System (FAERS) for drugs that are disproportionately reported with rhabdomyolysis.

**Methods:** We analyzed deduplicated FAERS reports from the first quarter of 2005 to the first quarter of 2025, considering only the primary suspect drug in each report. Signals were screened with the reporting odds ratio (ROR), requiring a lower 95% confidence limit above 1, more than 100 cases and a Bonferroni-adjusted *p* value below 0.01 (Fisher's exact test). Drugs that passed the screen were entered, together with age, sex and body weight, into a LASSO-penalized logistic regression, followed by an unpenalized model to estimate adjusted RORs (aRORs).

**Results:** We identified 36,225 rhabdomyolysis reports. Fifty-six drugs met the screening criteria, and 54 remained significant after adjustment ($p < 0.01$). The largest aRORs were for drugs with well-established muscle toxicity, including gemfibrozil (173.67), lovastatin (97.20) and simvastatin (85.12). Levetiracetam (aROR 11.02) and donepezil (aROR 8.90) were also disproportionately reported; both associations have been reported previously, and both estimates are susceptible to confounding by indication. About half of the reports with usable dates described onset within 30 days of starting the drug.

**Conclusions:** Rhabdomyolysis reports in FAERS involve a broad range of drugs beyond statins and fibrates. Because disproportionality measures do not estimate incidence, and the analysis did not account for indication, concomitant medication or overdose, these results are hypothesis-generating and should not be used to rank drugs by risk.

## 1 Introduction

Rhabdomyolysis is a clinical syndrome caused by the breakdown of skeletal muscle, with release of intracellular contents such as myoglobin and creatine kinase (CK) into the circulation [1, 2]. It ranges from asymptomatic CK elevation to severe illness complicated by acute kidney injury, electrolyte disturbances and, less often, disseminated intravascular coagulation [2–5]. Drugs and toxins are among its most common causes [6], and drug-associated rhabdomyolysis is of particular clinical interest because it is potentially preventable [7].

Statins are among the most widely prescribed drugs and the best-recognized cause of drug-associated muscle injury; the risk increases with higher doses and with pharmacokinetic interactions, such as those with gemfibrozil [8–10]. Many other drug classes have also been implicated, including antipsychotics, antidepressants and colchicine [6, 7], as well as antibiotics and antiseizure medications [11, 12]. Analyses of the FDA Adverse Event Reporting System (FAERS) have examined rhabdomyolysis reporting for individual drug classes, such as antibiotics, selective serotonin reuptake inhibitors, newer antiseizure medications, atypical antipsychotics and statins [11–15], and for specific drug combinations [16]. The Drug-Induced Rhabdomyolysis Atlas combined U.S. drug labeling with FAERS data for drugs whose labels mention rhabdomyolysis [17].

FAERS is a spontaneous reporting database that receives adverse event reports from healthcare professionals, consumers and manufacturers, and it is widely used for post-marketing signal detection [18]. Disproportionality analyses can identify drug–event combinations that are reported more often than expected, but they do not estimate incidence and are vulnerable to reporting biases and confounding [19, 20].

The aims of this study were (1) to screen FAERS for drugs disproportionately reported with rhabdomyolysis across all drug classes; (2) to estimate these associations with adjustment for age, sex and body weight; (3) to describe the characteristics of the reports and the reported time to onset; and (4) to explore whether drugs with signals mention rhabdomyolysis in their U.S. labels.

## 2 Methods

### *2.1 Data source*

Data were obtained from FAERS, which collects voluntary reports of adverse events from healthcare professionals, consumers and manufacturers worldwide [21]. All quarterly data files from the first quarter of 2005 to the first quarter of 2025 were extracted. Each quarterly release contains linked tables of demographic information (DEMO), adverse events (REAC), drugs (DRUG), outcomes (OUTC), report sources (RPSR), therapy dates (THER) and indications (INDI). All tables were imported into a PostgreSQL relational database.

Because a case can be updated and released again in later quarterly files, duplicate reports were removed: for reports with the same CASEID, the record with the most recent FDA_DT was retained, and when both CASEID and FDA_DT were identical, the record with the highest PRIMARYID was retained. Any remaining duplicate PRIMARYIDs were removed in a second step.

### *2.2 Case definition and drug exposure*

Rhabdomyolysis cases were identified in the REAC table using Medical Dictionary for Regulatory Activities (MedDRA) preferred terms corresponding to rhabdomyolysis [22]. Only the drug recorded as the primary suspect in each report was considered; secondary suspect and concomitant drugs were not analyzed. Reports lacking essential demographic or drug information were excluded.

### *2.3 Disproportionality analysis*

For each drug, a two-by-two contingency table was constructed comparing rhabdomyolysis reports with all other reports (Table 1). The ROR and its 95% confidence interval (CI) were calculated as follows [23, 24]:

$$\mathrm{ROR} = \frac{a/b}{c/d} = \frac{ad}{bc} \tag{1}$$

$$95\%\ \mathrm{CI} = \exp\left(\ln(\mathrm{ROR}) \pm 1.96\sqrt{\frac{1}{a} + \frac{1}{b} + \frac{1}{c} + \frac{1}{d}}\right) \tag{2}$$

A signal was defined as meeting all of the following criteria: (1) a lower limit of the 95% CI of the ROR greater than 1; (2) more than 100 rhabdomyolysis reports for the drug ($a > 100$); and (3) a Bonferroni-adjusted $p$ value below 0.01 from Fisher's exact test. The results were displayed as a volcano plot.

**Table 1.** Two-by-two contingency table used in the disproportionality analysis.

| | **Rhabdomyolysis reports** | **All other reports** |
|---|---|---|
| Drug of interest | *a* | *b* |
| All other drugs | *c* | *d* |

Counts refer to deduplicated reports; only the primary suspect drug of each report was considered.

### *2.4 Multivariable regression*

To adjust for age, sex and body weight, a logistic regression model was fitted in which the outcome was whether a report described rhabdomyolysis and the exposures were indicator variables for the drugs that met the screening criteria; reports with any other primary suspect drug formed the reference group. Reports with missing or implausible values were excluded (age ≤ 0 or ≥ 100 years, body weight ≤ 0 or ≥ 400 kg, or an invalid sex code). Predictors were first selected with LASSO-penalized logistic regression [25]. The penalty parameter (λ) was chosen by 10-fold cross-validation as the value that maximized the area under the receiver operating characteristic curve (AUC; $\lambda_{min}$). An unpenalized logistic regression including the demographic covariates and the drugs retained by LASSO was then fitted to the same data to obtain adjusted RORs (aRORs) with 95% CIs. Model discrimination was summarized with the AUC, and $p < 0.01$ was considered statistically significant.

### *2.5 Drug label review*

For the 54 drugs that remained significant in the regression analysis, the U.S. prescribing information available on DailyMed [26] was searched for rhabdomyolysis-related terms (“rhabdomyolysis”, “myopathy”, “myoglobinuria” and “creatine kinase elevation”). This was a keyword search that did not distinguish between label sections (for example, Warnings and Precautions versus Postmarketing Experience).

### *2.6 Descriptive and time-to-onset analyses*

The characteristics of rhabdomyolysis reports, including sex, age, outcomes, reporter type and reporting country, were summarized descriptively. For reports with usable therapy start and event dates, the time to onset was calculated as the interval between the start of the primary suspect drug and the onset of rhabdomyolysis, and was summarized with a violin and box plot and a cumulative distribution curve.

### *2.7 Software and ethics*

All data cleaning, statistical analyses and visualizations were performed in R (version 4.2.1), with PostgreSQL (version 15.0) used for data management. FAERS contains de-identified, publicly available data; ethics approval and informed consent were therefore not required.

## 3 Results

### *3.1 Characteristics of rhabdomyolysis reports*

After deduplication, 36,225 rhabdomyolysis reports were identified. Male patients accounted for 20,095 reports (55.5%) and female patients for 12,396 (34.2%); sex was not reported in the remainder. Patients aged 18–64 years accounted for 15,956 reports (44.0%) and those aged 65–85 years for 10,659 (29.4%) (Figure 1A). The annual number of reports increased over the study period, with the highest counts in 2018–2020 (Figure 1B).

Hospitalization was recorded in 18,238 reports (50.3%), a life-threatening event in 5,412 (14.9%) and death in 3,816 (10.5%) (Figure 1C); 36,007 reports (99.4%) met the criteria for a serious adverse event. Physicians submitted 14,096 reports (38.9%). The most frequent reporting countries were the United States (11,220 reports; 31.0%), France (4,311; 11.9%) and Japan (4,079; 11.3%).

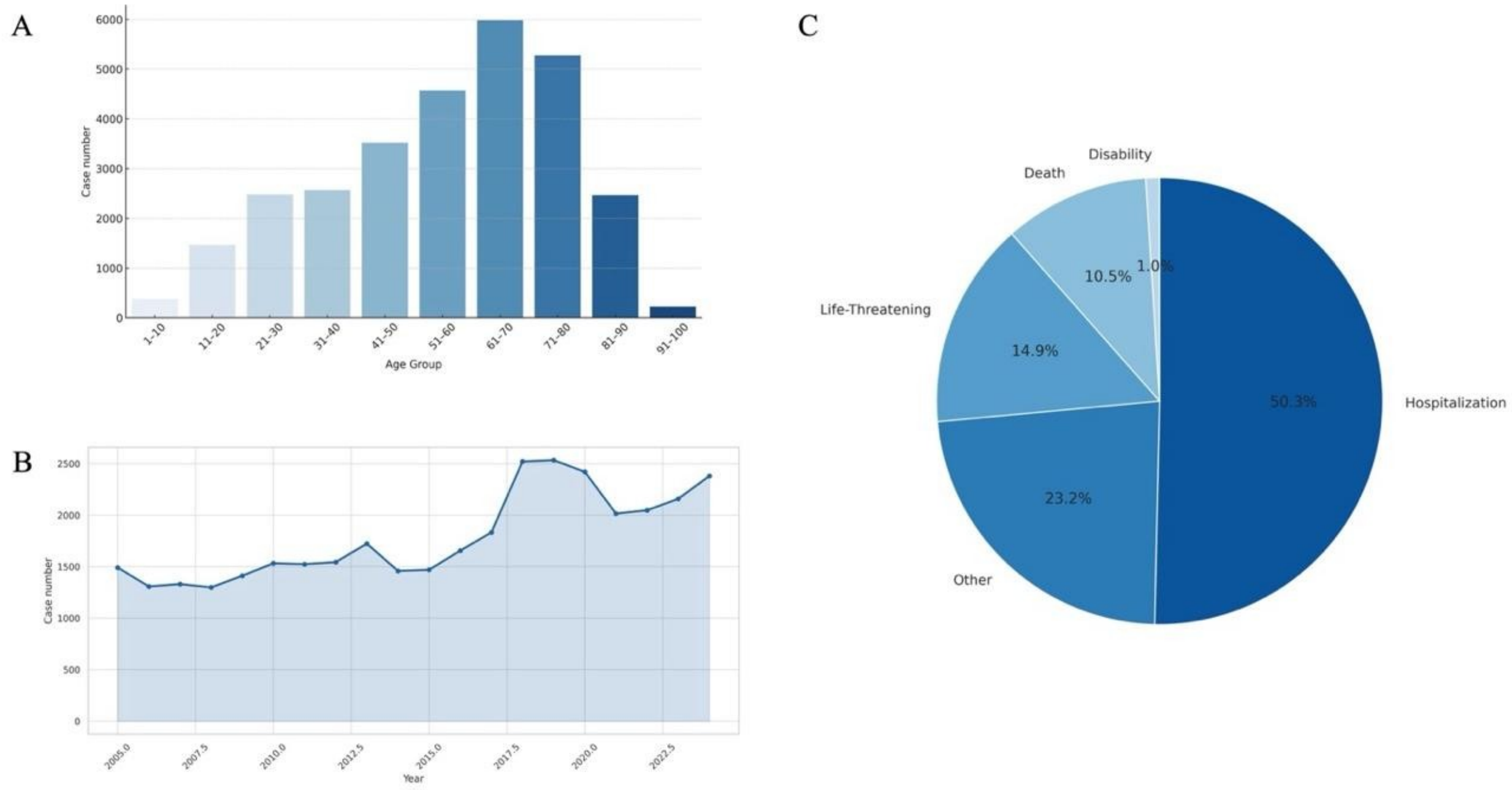

**Figure 1.** Characteristics of rhabdomyolysis reports. (A) Age distribution of reports with known age. (B) Annual number of reports, 2005–2024. (C) Distribution of reported outcomes.

### *3.2 Disproportionality screen*

Fifty-six drugs met all three signal criteria (Figure 2). Because *p* values in a database of this size depend largely on the number of reports, the most extreme *p* values were observed for frequently reported drugs, including statins (simvastatin, atorvastatin and rosuvastatin), levetiracetam, quetiapine and propofol.

### *3.3 Adjusted analysis*

At $\lambda_{min}$, LASSO retained all 56 drug indicators and the demographic covariates (Figure 3A, B). In the subsequent unpenalized model, 54 of the 56 drugs had aRORs significantly greater than 1 ($p < 0.01$). For descriptive purposes, these drugs were grouped by the size of the aROR: 14 drugs had an aROR ≥ 10, 18 had an aROR of 5 to < 10, and 22 had an aROR < 5. These groups describe the strength of disproportionate reporting only and should not be interpreted as levels of clinical risk.

The largest aRORs were observed for gemfibrozil (173.67), lovastatin (97.20) and simvastatin (85.12), and high estimates were also seen for the ezetimibe/simvastatin combination, other lipid-lowering drugs, daptomycin, propofol and colchicine (Figure 3D). The aROR was 11.02 for levetiracetam and 8.90 for donepezil.

Reports concerning male patients had higher adjusted odds of describing rhabdomyolysis (aROR 2.16; $p < 0.001$). The association with age was non-linear, with the highest estimate in the 20–30-year age group (aROR 1.44; $p < 0.001$). Body weight had a statistically significant but negligible association (aROR 0.999 per kg; $p = 0.004$). The AUC of the final model was 0.804 (Figure 3C); because it was estimated in the same data used to fit the model, this value is likely to be optimistic.

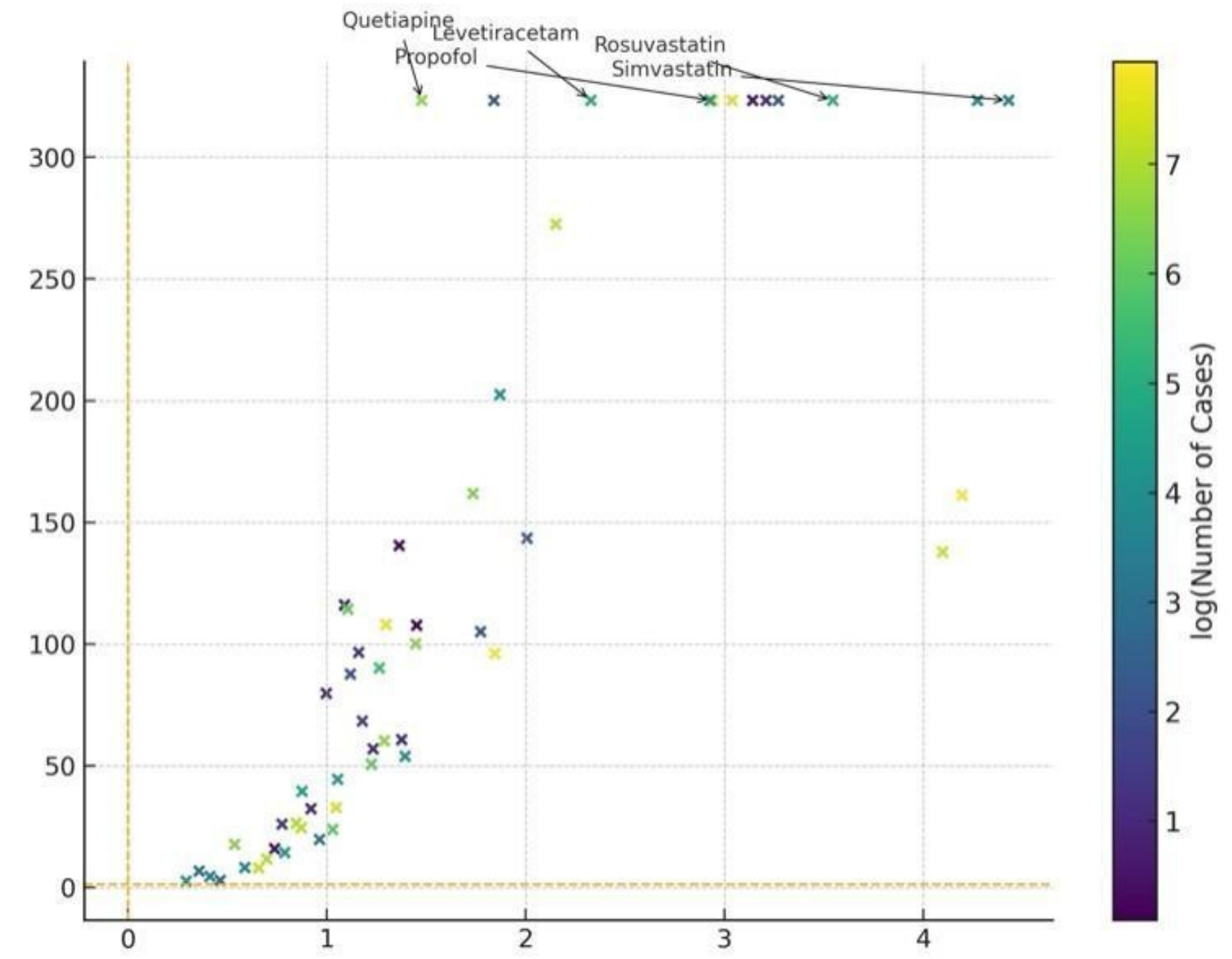

**Figure 2.** Volcano plot of the disproportionality screen. Each point represents one drug. The x-axis shows the natural logarithm of the ROR, and the y-axis shows $-\log_{10}$ of the Bonferroni-adjusted *p* value; color indicates the logarithm of the number of rhabdomyolysis reports. Points at the top of the plot correspond to *p* values below the numerical limit of the software (about $10^{-323}$) and are plotted at that limit. Selected drugs are labeled.

### *3.4 Time to onset*

Among reports with usable dates, the distribution of time to onset was strongly right-skewed (Figure 4A). About half of these reports described onset within the first 30 days of treatment, and the cumulative proportion increased more slowly thereafter (Figure 4B). A minority of reports described onset more than a year after the drug was started.

### *3.5 Drug label review*

In the keyword review, 33 of the 54 drugs (61.1%) had no mention of rhabdomyolysis-related terms in their U.S. labels. This result should be interpreted with caution. During revision of this manuscript, the current U.S. label for donepezil was found to list rhabdomyolysis among postmarketing adverse reactions [27]; at least this drug was therefore misclassified, and the proportion is likely to be an overestimate.

## 4 Discussion

### *4.1 Principal findings*

In this analysis of two decades of FAERS data, 56 drugs met predefined criteria for disproportionate reporting of rhabdomyolysis, and 54 remained significant after adjustment for age, sex and body weight. The largest estimates were mostly for drugs with well-established muscle toxicity, such as fibrates, statins, daptomycin, colchicine and propofol, but the drugs with signals also included antipsychotics, antidepressants, benzodiazepines, antiseizure medications and donepezil (Figure 3D). Many of these drugs have been identified in earlier FAERS analyses of individual drug classes [11–15].

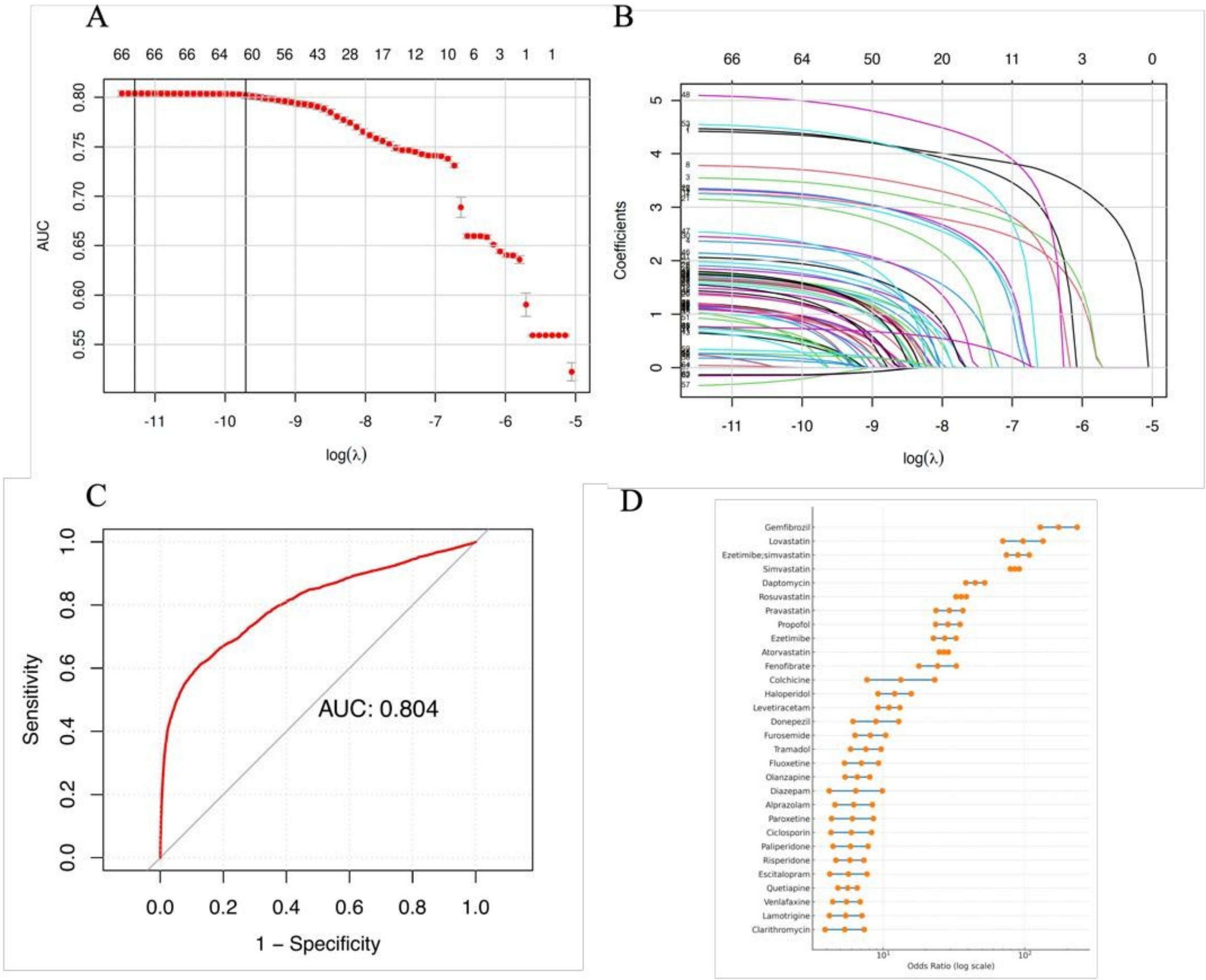

**Figure 3.** Adjusted analysis. (A) Cross-validated AUC across values of the LASSO penalty parameter, log(λ). The numbers above the plot are the numbers of non-zero coefficients, and the vertical lines mark $\lambda_{min}$ (left) and $\lambda_{1se}$ (right). (B) LASSO coefficient paths. (C) Receiver operating characteristic curve of the final model (AUC 0.804). (D) Adjusted RORs with 95% CIs (log scale) for the 30 drugs with the largest estimates.

These findings should be interpreted as signals of disproportionate reporting rather than as estimates of risk. FAERS contains no information on the number of patients exposed to each drug, so neither RORs nor aRORs measure incidence, and their size is not a reliable guide to the relative risk of different drugs [19, 20].

### *4.2 Known associations*

Lovastatin and simvastatin had markedly higher estimates than rosuvastatin, pravastatin and atorvastatin (Figure 3D). This pattern is compatible with known pharmacology: lovastatin and simvastatin are metabolized by CYP3A4 and are particularly susceptible to interactions that increase statin exposure [10]. The very large aROR for gemfibrozil is consistent with its inhibition of statin elimination [10] and with earlier FAERS data showing much higher reporting rates of rhabdomyolysis for gemfibrozil–statin than for fenofibrate–statin combinations [28]. Because only primary suspect drugs were analyzed, however, the estimate for gemfibrozil probably reflects concomitant statin use to a substantial extent; the high estimate for ezetimibe, which is often prescribed together with a statin, may reflect the same problem. Reporting of statin-associated muscle injury may also have been increased by long-standing regulatory and clinical attention (notoriety bias) [20, 29]. The agreement of these estimates with known pharmacology is reassuring, but it does not show that the method can quantify risk.

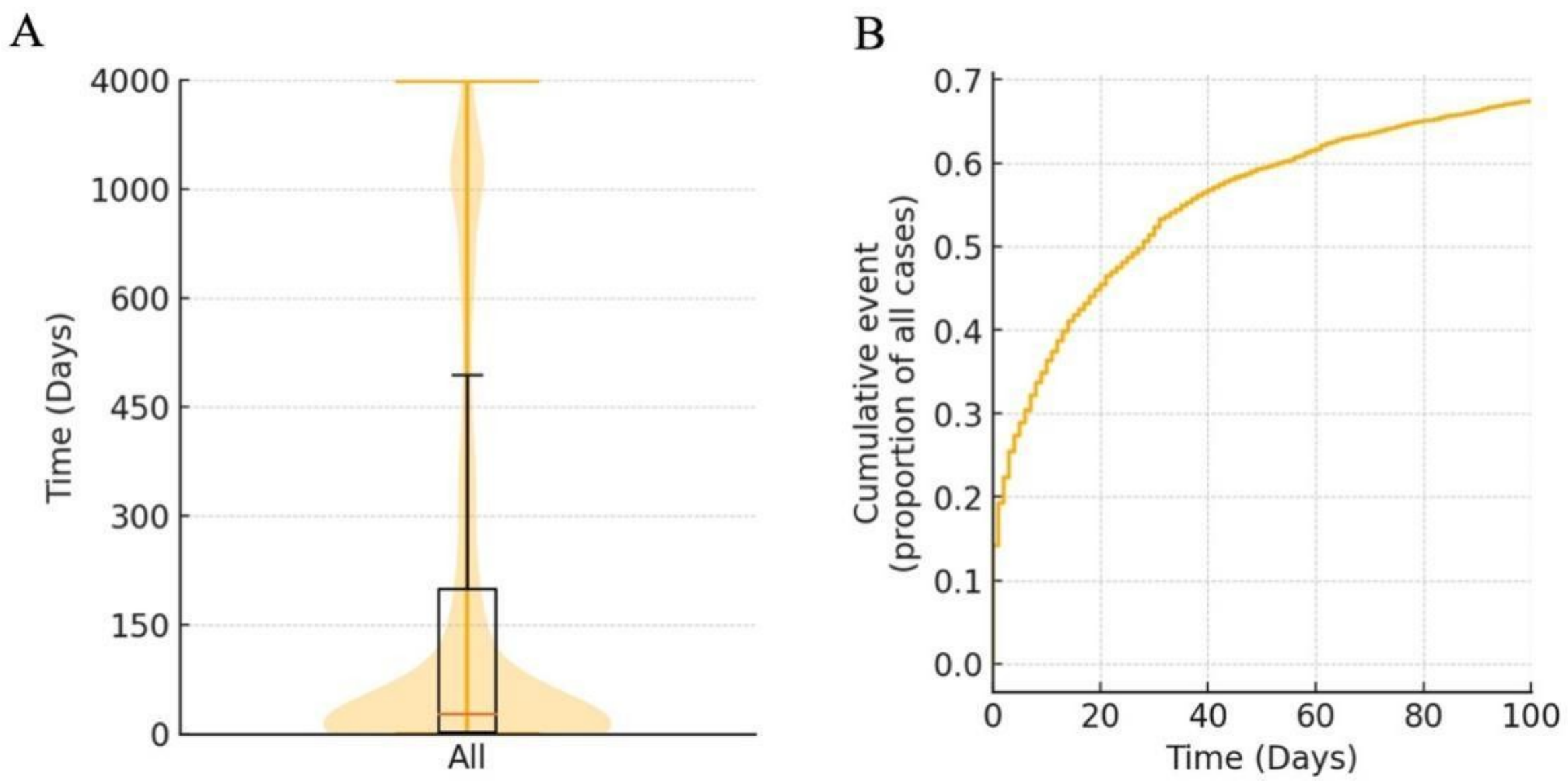

**Figure 4.** Reported time to onset of rhabdomyolysis after the start of the primary suspect drug (all drugs combined). (A) Violin and box plot; values were winsorized at the 1st and 99th percentiles and outliers are not shown. Note the non-linear y-axis. (B) Cumulative proportion of reports by time to onset during the first 100 days.

### *4.3 Levetiracetam and donepezil*

*Levetiracetam.* Rhabdomyolysis during levetiracetam treatment has been described in a series of case reports, 13 of which were summarized in a 2020 review [30], and an earlier FAERS analysis of newer antiseizure medications found the highest ROR for levetiracetam (8.01) [12]. Regulators in Japan (2015), the European Union (2016) and Canada (2017) have added rhabdomyolysis to levetiracetam product information [31–33], whereas the current U.S. label does not mention it [34]. Our finding therefore adds to existing evidence rather than representing a new signal, and its magnitude should be interpreted with particular caution. Seizures, especially generalized convulsive seizures and status epilepticus, can themselves cause rhabdomyolysis [1, 2], and levetiracetam is widely used in the acute treatment of seizures, including established status epilepticus [35]. Confounding by indication and protopathic bias are therefore likely, and we did not adjust for indication or exclude reports with co-reported seizures. Regulatory attention in Japan may also have stimulated reporting there; Health Canada has noted that patients of Japanese origin may be at higher risk [33]. If the association is causal, its mechanism is unknown. Synaptic vesicle protein 2A (SV2A), the binding site of levetiracetam [36], is present in mouse motor nerve terminals that innervate slow-twitch muscle fibers [37], but there is no direct evidence that this is relevant to muscle injury. Comparisons restricted to patients treated with other antiseizure medications, and exclusion of reports with co-reported seizures, would help to separate the effect of the drug from that of the underlying disease.

*Donepezil.* Health Canada added warnings about rhabdomyolysis and neuroleptic malignant syndrome to donepezil labeling in 2015 [38], and the current U.S. label lists rhabdomyolysis among postmarketing adverse reactions [27]. In a population-based cohort of adults aged 66 years or older in Ontario, Canada, the 30-day risk of hospitalization with rhabdomyolysis was higher after starting donepezil than after starting rivastigmine or galantamine (0.06% vs 0.02%; weighted odds ratio 2.21, 95% CI 1.52–3.22) [39]. Our aROR of 8.90 cannot be compared directly with this estimate: it compares reporting with all other drugs in FAERS rather than incidence with other cholinesterase inhibitors, and it is likely to be inflated by confounding, because patients with dementia are older and frailer and are prone to falls and prolonged immobilization, which are themselves causes of rhabdomyolysis [2]. A peripheral mechanism is plausible, as donepezil inhibits acetylcholinesterase at the neuromuscular junction and prolongs the action of acetylcholine in isolated mouse muscle [40], but its role in rhabdomyolysis remains to be established.

### *4.4 Signals and drug labeling*

Many drugs with signals did not mention rhabdomyolysis in their U.S. labels, but this does not by itself indicate a labeling deficiency. A disproportionality signal requires assessment of individual case reports, clinical plausibility and evidence from other data sources before it can support a change in labeling [19, 20]. The case of levetiracetam illustrates that regulators may reach different conclusions about the same signal [31–34]. A more informative comparison would require a structured review of label sections and versions, together with the evidence supporting each signal.

### *4.5 Limitations*

This study has several important limitations. First, FAERS contains no information on the number of patients exposed, so RORs and aRORs cannot be interpreted as measures of risk, and the grouping of drugs by aROR should not be used for clinical risk stratification. Spontaneous reports are also affected by under-reporting, duplicate and stimulated reporting, and notoriety bias [20, 29], and the proportions of reports with serious outcomes, including death, are not case-fatality rates.

Second, the analysis did not account for indication or underlying disease. Confounding by indication is likely for several drugs, including levetiracetam (seizures), donepezil (frailty, falls and immobilization) and propofol (critical illness).

Third, only the primary suspect drug was analyzed, so concomitant medications and drug–drug interactions were ignored. This is an important limitation for rhabdomyolysis, in which interactions with statins, for example involving gemfibrozil, clarithromycin, colchicine or ciclosporin, play a major role [10, 16]. Using all other drugs as the comparator may also have introduced competition bias: frequently reported drugs such as statins raise the background frequency of rhabdomyolysis reports and can mask signals for other drugs [41].

Fourth, reports of overdose or intentional misuse were not excluded. Toxic exposures, including drug overdose, are a common cause of rhabdomyolysis [6], partly through prolonged immobilization [2], and this may have contributed to the signals for drugs such as benzodiazepines, antidepressants, antipsychotics and tramadol.

Fifth, cases were identified by preferred-term matching rather than a Standardised MedDRA Query, the exact terms and MedDRA version are not reported here, and individual reports were not reviewed to confirm the diagnosis or assess causality. Combination products, such as ezetimibe/simvastatin, were treated as separate entities.

Sixth, the statistical approach has limitations. Drugs were screened and then modeled in the same data, so the $p$ values and CIs of the final model do not account for the selection step; the $p$ values are likely to be too small and the CIs too narrow [42]. LASSO retained all screened drugs and therefore did not reduce the model, and the AUC was estimated in the training data. The regression was restricted to reports with complete and plausible age, sex and body weight, which are often missing in FAERS; the number of reports included in the regression is not reported here, and this subset may not be representative of all reports. Finally, with very large numbers of reports, $p$ values mainly reflect the volume of reporting and are of limited value for judging the importance of a signal.

Seventh, the time-to-onset analysis pooled all drugs and was restricted to reports with usable dates; it describes the timing of reported events, not how risk changes over time. Eighth, the label review was a keyword search that did not distinguish label sections or versions, and it misclassified at least one drug.

Lastly, this report does not include all items recommended by the READUS-PV guideline for reporting disproportionality analyses [43], which was developed partly in response to frequent shortcomings in published studies [44]. The signals identified here should be evaluated with designs that address these problems, such as disproportionality analyses restricted to active comparators with the same indication, case-level review, and studies in data sources with information on exposed populations, such as claims or electronic health records, using active-comparator, new-user designs [45].

## 5 Conclusions

Rhabdomyolysis reports in FAERS involve a broad range of drugs. The strongest signals were for drugs with well-established muscle toxicity, whereas the signals for drugs such as levetiracetam and donepezil are consistent with previous reports but are likely to be affected by confounding. Because disproportionality analysis cannot estimate risk, these findings should be regarded as hypothesis-generating and require confirmation in studies designed to estimate risk.

## Declarations

**Acknowledgements.** The author thanks the U.S. Food and Drug Administration for making FAERS data publicly available. The findings and interpretations are the author's own and do not necessarily represent the views of the FDA.

**Author contributions.** Enpu Liang is the sole author and was responsible for all aspects of the work.

**Funding.** No funding was received for this work.

**Data availability.** FAERS data are publicly available from the U.S. Food and Drug Administration [21].

**Note on version 2.** Version 2 corrects and updates the reference list, revises the interpretation of the results (disproportionality estimates are presented as measures of reporting association rather than of risk), corrects statements about the novelty and labeling status of the levetiracetam and donepezil signals, and adds a discussion of limitations. The data and numerical results are unchanged from version 1.

## References

1. Torres PA, Helmstetter JA, Kaye AM, Kaye AD. Rhabdomyolysis: pathogenesis, diagnosis, and treatment. *Ochsner J.* 2015;15(1):58-69.
2. Bosch X, Poch E, Grau JM. Rhabdomyolysis and acute kidney injury. *N Engl J Med.* 2009;361(1):62-72.
3. Chavez LO, Leon M, Einav S, Varon J. Beyond muscle destruction: a systematic review of rhabdomyolysis for clinical practice. *Crit Care.* 2016;20(1):135.
4. Sauret JM, Marinides G, Wang GK. Rhabdomyolysis. *Am Fam Physician.* 2002;65(5):907-912.
5. Khan FY. Rhabdomyolysis: a review of the literature. *Neth J Med.* 2009;67(9):272-283.
6. Melli G, Chaudhry V, Cornblath DR. Rhabdomyolysis: an evaluation of 475 hospitalized patients. *Medicine (Baltimore).* 2005;84(6):377-385.
7. Hohenegger M. Drug induced rhabdomyolysis. *Curr Opin Pharmacol.* 2012;12(3):335-339.
8. Ward NC, Watts GF, Eckel RH. Statin toxicity. *Circ Res.* 2019;124(2):328-350.

9. Chatzizisis YS, Koskinas KC, Misirli G, Vaklavas C, Hatzitolios A, Giannoglou GD. Risk factors and drug interactions predisposing to statin-induced myopathy: implications for risk assessment, prevention and treatment. *Drug Saf.* 2010;33(3):171-187.
10. Neuvonen PJ, Niemi M, Backman JT. Drug interactions with lipid-lowering drugs: mechanisms and clinical relevance. *Clin Pharmacol Ther.* 2006;80(6):565-581.
11. Teng C, Baus C, Wilson JP, Frei CR. Rhabdomyolysis associations with antibiotics: a pharmacovigilance study of the FDA Adverse Event Reporting System (FAERS). *Int J Med Sci.* 2019;16(11):1504-1509.
12. Deng Z, Wang S, Wu C. Rhabdomyolysis associated with newer-generation anti-seizure medications (ASMs): a real-world retrospective and pharmacovigilance study. *Front Pharmacol.* 2023;14:1197470.
13. Wang Y, Lin Y, Lin Q, Liang H, Cai W, Jiang D. Exploring the association between selective serotonin reuptake inhibitors and rhabdomyolysis risk based on the FDA pharmacovigilance database. *Sci Rep.* 2023;13:12257.
14. Yin Y, Jiang J, Jin Y. Evaluation of atypical antipsychotics associated rhabdomyolysis using the FDA adverse event reporting system database. *Sci Rep.* 2025;15:10499.
15. Morris R, Bu K, Han W, et al. The association between statin drugs and rhabdomyolysis: an analysis of FDA Adverse Event Reporting System (FAERS) data and transcriptomic profiles. *Genes (Basel).* 2025;16(3):248.
16. Zhang S, Yan MM, Zhao H, Qiu XY, Zhu D. Rhabdomyolysis associated with concomitant use of colchicine and statins in the real world: identifying the likelihood of drug–drug interactions through the FDA adverse event reporting system. *Front Pharmacol.* 2024;15:1445324.
17. Wen Z, Liang Y, Hao Y, et al. Drug-induced rhabdomyolysis atlas (DIRA) for idiosyncratic adverse drug reaction management. *Drug Discov Today.* 2019;24(1):9-15.
18. Sakaeda T, Tamon A, Kadoyama K, Okuno Y. Data mining of the public version of the FDA Adverse Event Reporting System. *Int J Med Sci.* 2013;10(7):796-803.
19. Bate A, Evans SJW. Quantitative signal detection using spontaneous ADR reporting. *Pharmacoepidemiol Drug Saf.* 2009;18(6):427-436.
20. Fusaroli M, Sartori D, van Puijenbroek EP, Norén GN. Charting and sidestepping the pitfalls of disproportionality analysis. *Drug Saf.* 2026;49(2):147-162.
21. U.S. Food and Drug Administration. FDA Adverse Event Reporting System (FAERS): latest quarterly data files. https://www.fda.gov/drugs/fdas-adverse-event-reporting-system-faers/fda-adverse-event-reporting-system-faers-latest-quarterly-data-files
22. MedDRA Maintenance and Support Services Organization. Medical Dictionary for Regulatory Activities (MedDRA). https://www.meddra.org/
23. Rothman KJ, Lanes S, Sacks ST. The reporting odds ratio and its advantages over the proportional reporting ratio. *Pharmacoepidemiol Drug Saf.* 2004;13(8):519-523.
24. van Puijenbroek EP, Bate A, Leufkens HGM, Lindquist M, Orre R, Egberts ACG. A comparison of measures of disproportionality for signal detection in spontaneous reporting systems for adverse drug reactions. *Pharmacoepidemiol Drug Saf.* 2002;11(1):3-10.
25. Tibshirani R. Regression shrinkage and selection via the lasso. *J R Stat Soc Series B Stat Methodol.* 1996;58(1):267-288.
26. U.S. National Library of Medicine. DailyMed. https://dailymed.nlm.nih.gov/dailymed/
27. Eisai Inc. ARICEPT (donepezil hydrochloride) tablets and orally disintegrating tablets: prescribing information. Revised December 2021. DailyMed set ID 98e451e1-e4d7-4439-a675-c5457ba20975. https://dailymed.nlm.nih.gov/dailymed/drugInfo.cfm?setid=98e451e1-e4d7-4439-a675-c5457ba20975
28. Jones PH, Davidson MH. Reporting rate of rhabdomyolysis with fenofibrate + statin versus gemfibrozil + any statin. *Am J Cardiol.* 2005;95(1):120-122.
29. Moore N, Hall G, Sturkenboom M, Mann R, Lagnaoui R, Begaud B. Biases affecting the proportional reporting ratio (PRR) in spontaneous reports pharmacovigilance databases: the example of sertindole. *Pharmacoepidemiol Drug Saf.* 2003;12(4):271-281.

30. Moinuddin IA. Suspected levetiracetam-induced rhabdomyolysis: a case report and literature review. *Am J Case Rep.* 2020;21:e926064.
31. Pharmaceuticals and Medical Devices Agency. Summary of investigation results: levetiracetam (revision of precautions: rhabdomyolysis). January 9, 2015. https://www.pmda.go.jp/files/000197807.pdf
32. European Medicines Agency. Keppra (levetiracetam): scientific conclusions and grounds for the variation to the terms of the marketing authorisation, PSUSA/00001846/201511. September 2016. https://www.ema.europa.eu/en/documents/scientific-conclusion/keppra-h-c-psusa-1846-201511-epar-scientific-conclusion_en.pdf
33. Health Canada. Health Product InfoWatch: September 2017 (product monograph update: Keppra [levetiracetam], rhabdomyolysis). September 2017. https://www.canada.ca/en/health-canada/services/drugs-health-products/medeffect-canada/health-product-infowatch/health-product-infowatch-september-2017/page-2.html
34. UCB Inc. KEPPRA (levetiracetam) tablets and oral solution: prescribing information. Revised March 2024. DailyMed set ID 3ca9df05-a506-4ec8-a4fe-320f1219ab21. https://dailymed.nlm.nih.gov/dailymed/drugInfo.cfm?setid=3ca9df05-a506-4ec8-a4fe-320f1219ab21
35. Kapur J, Elm J, Chamberlain JM, et al. Randomized trial of three anticonvulsant medications for status epilepticus. *N Engl J Med.* 2019;381(22):2103-2113.
36. Lynch BA, Lambeng N, Nocka K, et al. The synaptic vesicle protein SV2A is the binding site for the antiepileptic drug levetiracetam. *Proc Natl Acad Sci U S A.* 2004;101(26):9861-9866.
37. Chakkalakal JV, Nishimune H, Ruas JL, Spiegelman BM, Sanes JR. Retrograde influence of muscle fibers on their innervation revealed by a novel marker for slow motoneurons. *Development.* 2010;137(20):3489-3499.
38. Health Canada. Information update: Alzheimer’s drug Aricept (donepezil) – new warnings on the serious risks of muscle breakdown and of a neurological disorder. January 21, 2015. https://recalls-rappels.canada.ca/en/alert-recall/alzheimer-s-drug-aricept-donepezil-new-warnings-serious-risks-muscle-breakdown-and
39. Fleet JL, McArthur E, Patel A, Weir MA, Montero-Odasso M, Garg AX. Risk of rhabdomyolysis with donepezil compared with rivastigmine or galantamine: a population-based cohort study. *CMAJ.* 2019;191(37):E1018-E1024.
40. Redman RR, Mackenzie H, Dissanayake KN, Eddleston M, Ribchester RR. Donepezil inhibits neuromuscular junctional acetylcholinesterase and enhances synaptic transmission and function in isolated skeletal muscle. *Br J Pharmacol.* 2022;179(24):5273-5289.
41. Pariente A, Didailler M, Avillach P, et al. A potential competition bias in the detection of safety signals from spontaneous reporting databases. *Pharmacoepidemiol Drug Saf.* 2010;19(11):1166-1171.
42. Berk R, Brown L, Buja A, Zhang K, Zhao L. Valid post-selection inference. *Ann Stat.* 2013;41(2):802-837.
43. Fusaroli M, Salvo F, Begaud B, et al. The reporting of a disproportionality analysis for drug safety signal detection using individual case safety reports in PharmacoVigilance (READUS-PV): development and statement. *Drug Saf.* 2024;47(6):575-584.
44. Khouri C, Revol B, Lepelley M, et al. A meta-epidemiological study found lack of transparency and poor reporting of disproportionality analyses for signal detection in pharmacovigilance databases. *J Clin Epidemiol.* 2021;139:191-198.
45. Lund JL, Richardson DB, Stürmer T. The active comparator, new user study design in pharmacoepidemiology: historical foundations and contemporary application. *Curr Epidemiol Rep.* 2015;2(4):221-228.